\documentclass{aastex}
\usepackage{spr-astr-addons}
\usepackage{url}

\RequirePackage{color}

\begin{document}

\title{Determination of Intrinsic Mode and Linear Mode Coupling
in Solar Microwave Bursts}

\shortauthors{Huang et al.} \shorttitle{Intrinsic Mode and Linear
Mode Coupling}

\author{Guangli Huang$^{1,2}$, Qiwu Song$^{1,2}$, and Jianping Li$^{1,2}$}

\affil{$^1$Purple Mountain Observatory, Chinese Academy of Sciences
(CAS), Nanjing, 210008, China\\$^2$Key Laboratory of Dark Matter and
Space Astronomy, CAS, Nanjing, 210008, China}

\begin{abstract}
An explicit equation of the propagational angle of microwave
emission between the line-of-sight and the local magnetic field is
newly derived based on the approximated formulae of nonthermal
gyrosynchrotron emission (Dulk and Marsh 1982). The existence of the
solution of propagational angle is clearly shown under a series of
typical parameters in solar microwave observations. It could be used
to determine the intrinsic mode and linear mode coupling in solar
microwave bursts by three steps. 1) The mode coupling may happen
only when the angle approximately equals to 90 degrees, i.e., when
the emission propagates through the quasi-transverse region (Cohen
1960). 2) The inversion of polarization sense due to the weakly mode
coupling takes place only when the transition frequency defined by
Cohen (1960) is larger than the frequency of microwave emission, and
an observable criterion of the weakly mode coupling in flaring loops
was indicated by the same polarization sense in the two footpoints
of a flaring loop (Melrose and and Robinson 1994). 3) Finally, the
intrinsic mode of microwave emission is determined by the observed
polarization and the calculated direction of local magnetic field
according to the general plasma dispersion relation, together with
the mode coupling process. However, a 180-degree ambiguity still
exists in the direction of longitudinal magnetic field, to produce
an uncertainty of the intrinsic mode. One example is selected to
check the feasibility of the method in the 2001 September 25 event
with a loop-like structure nearby the central meridian passage
observed by Nobeyama Radio Heliograph and Polarimeters. The
calculated angle in one footpoint (FP) varied around $90^{\circ}$ in
two time intervals of the maximum phase, which gives a direct
evidence of the emission propagating through a quasi-transverse
region where the linear mode coupling took place, while, the angle
in another FP was always smaller than $90^{\circ}$ where the mode
coupling did not happen. Moreover, the right-circular sense at 17
GHz was always observed in both two FPs during the event, which
supports that the transition frequency should be larger than 17 GHz
in the first FP together with strong magnetic field of over 2000
Gauses in photosphere, where the weakly coupled case should happen.
Moreover, there are two possibilities of the intrinsic mode in the
two FPs due to the 180-degree ambiguity. 1) The emission of
extraordinary (X) mode from the first FP turns to the ordinary (O)
mode in the two time intervals of the maximum phase, while, the
X-mode is always emitted from the second FP. 2) The inversion from
O-mode to X-mode takes place in the first FP, while the O-mode keeps
in the second FP. If the magnetic polarities in photosphere and
corona are coincident in this event, the intrinsic mode belongs to
the second case.
\end{abstract}


\section{INTRODUCTION}

Radio polarization is generally considered as an important parameter
for measuring the local magnetic field, as well as for determining
the intrinsic mode of electromagnetic emissions from the local
plasmas. The spatial resolvable data of Stokes I and V components
are provided by a series of solar radio instruments in different
frequencies, such as Nobeyama Radioheliograph (NoRH), Nancy Radio
Heliograph (NRH), Siberian Solar Radio Telescope (SSRT), Owens
Valley Solar Arrays (OVSA), and etc. Especially, high-quality images
of Stokes I and V components at 17 GHz are continuously recorded
from 1992 up to date in NoRH, and intensively studied by solar
astronomers in all over the world. But only limited authors paid
particular attention to the radio polarization observed by NoRH.

In five events (Kundu {\it et al.} 2001), the NoRH images show
direct evidence that the radio sources are compact bipolar loops:
source sizes are less than 5", and three of the five events studied
show closely spaced oppositely polarized components in the circular
polarization maps. All five events are located directly over
magnetic neutral lines in the photosphere, which is consistent with
a single-loop scenario in which magnetic energy release and
acceleration of nonthermal electrons are confined to a compact
localized region. The polarization is analyzed in four microwave
bursts observed by NoRH with one looptop (LT) and two footpoint (FP)
sources (Su and Huang 2004). The three sources in each given burst
are always polarized in the same sense. This may be interpreted in
terms of extraordinary mode (X-mode) emission, by taking into
account the polarity of the underlying magnetic field and
propagation effects, which may lead to inversion of the sense of
polarization in the limbward FP and LT sources of the flaring loop.
Huang and Lin (2006) found a quasi-periodic reversal between
left-circular polarization (LCP) and right-circular polarization
(RCP) in the LT source, five minutes before the 2002 April 21 flare
observed by NoRH, and the polarization gradually turned to LCP.
During this period, the polarization of the corresponding FP source
maintained the RCP sense. It may be taken as evidence that magnetic
energy is released or energetic particles are produced at the
magnetic reconnection site in a quasi-periodic fashion.

Moreover, in the 2001 March 10 flare observed simultaneously by NoRH
and SSRT, the emission at 17 GHz with LCP is generated in positive
magnetic field polarity of the Michelson Doppler Imager (MDI)
photospheric magnetogram. It is considered as an evidence of the
ordinary mode (O-mode) emission and explained by strong anisotropy
of the pitch angle distribution of the nonthermal electrons that
responsible for the radio emission (Altyntzev {\it et al.} 2008).
In a recent statistics of Huang {\it et al.} (2010), two footpoint
emissions are compared in a total of 24 NoRH events with loop-like
structures. The polarization degrees in the two FPs are well
correlated, while, they mostly have the same sense (RCP or LCP).
There is a comparable proportion of the normal and abnormal events,
which are defined as whether a stronger emission corresponds to
a stronger or weaker polarization in the two FP sources
(Kundu {\it et al.} 1995).

On the other hand, the observed polarization sense (RCP or LCP)
may be reversed due to the linear mode coupling mechanism (Cohen
1960; Melrose and Robinson 1994; Zheleznyakov {\it et al.} 1996),
on the basis of plasma dispersion relation under approximations of
locally homogeneous corona and geometrical optics, which is widely
used to analyze the polarization of solar radio emission (e.g.,
Peterova and Akhmedov 1974). When the emission propagating along
the line-of-sight is nearly perpendicular to the local magnetic
field (QT-region), the observed polarization sense is reversed
in those frequencies below the transition frequency (weakly coupled),
defined by Cohen (1960). Otherwise, the sense is unchanged in those
frequencies above the transition frequency (strongly coupled).
For two footpoint emissions with opposite magnetic polarities,
they should show the same polarization sense in the weakly coupled
case, but opposite polarization senses in the strongly coupled case
(Melrose and Robinson 1994). However, it is not easy to judge if
the observed emission does actually propagate through the QT-region,
and the observations were not always compatible with that predicted
by the theory (Mclean and Sheridan 1972; White {\it et al.} 1992).

It is predicted by Ramaty (1969) for the nonthermal gyrosynchrotron
(GS) emission that the extraordinary mode (X-mode) is dominated in
an optically-thin source, and the O-mode is only dominated in an
optically-thick source. One case to observe the optically-thin
O-mode emission in microwave band is caused by the thermal
gyroresonance absorption at the third harmonic (Preka-Papadema and
Alissandrakis 1988), the suppression of X-mode emission, originating
in the lower layers of the loop, will produce excess O-mode
radiation at the diskward foot of a flaring loop. This requires that
the magnetic field is strong enough to bring the third harmonic
layer inside the flaring loop. Thus, Alissandrakis {\it et al.}
(1993) have found evidence for the O-mode emission in two classes of
events. In one class the O-mode comes from the regions overlying the
strong magnetic field, which can be interpreted in terms of the
thermal gyroresonance absorption at the third harmonic. In the other
class the entire burst emits in the O-mode, which may be attributed
to high GS optical depth. The second case of optically-thin O-mode
emission in microwave band was reported by Ledenev {\it et al.}
(2002) in the 2000 July 14 event, due to the scattering of
electromagnetic radiation during propagation through a plasma layer
with developed Langmuir turbulence. The ordinary component is
slightly lowered, while the extraordinary component undergoes the
most effective scattering. The third case of optically-thin O-mode
emission in microwave band was suggested by White et al. (1992) and
Gopalswamy {\it et al.} (1994), and the linear mode coupling may
take place in the plasma current sheet with a transverse component
of magnetic field. It was also reported by Vourlidas {\it et al.}
(1997) that the O-mode emission in 4.7 GHz observed by VLA in a
solar active region, in contradiction with the gyroresonance models.
The excess of the O-mode emission is attributed to the magnetic
field configuration and the temperature inhomogeneities across the
spot, and it may originate from the hotter penumbral loops.

Therefore, it is convenient to study the linear mode coupling and
the intrinsic mode of electromagnetic emissions from the observed
polarization sense and photospheric magnetic polarity. However, it
is difficult to measure the local magnetic field of radio sources
with lower plasma density in solar corona than that in the
photosphere. For example, the radio emission at 17 GHz is generated
at least ten thousand kilometers above the photosphere, and the
photospheric field lines may turn to different directions in solar
corona. Section 2 shows one method to calculate the propagational
angle of microwave emission between the line-of-sight and the local
magnetic field, thus, to determine the linear mode coupling process
and the intrinsic mode in solar microwave bursts. Section 3 analyzes
one example observed by NoRH. Relevant discussions are given in
Section 4.

\section{METHOD}

There are two equations derived in Huang (2006) for the coronal
magnetic field strength ($B$), the angle ($\theta$) between the the
coronal magnetic field and the line-of-sight (i.e., Eqs.(8) and (9)
in Huang 2006), based on the nonthermal GS formulae in Dulk and Marsh
(1982), with an error of about 30\% in respect to the full expressions
of nonthermal GS emission, for a limited range of electron spectral
index, harmonic number, propagation angle etc. The detailed derivations
and discussions are given in Huang (2006), and followed an earlier paper
of Zhou and Karlick\'y (1994).

\begin{equation}
\log\left({\frac{\nu}{\nu_B}}\right)={\frac{A_1+0.5A_2\log(1-x^2)}{A_3}},
\end{equation}

\begin{equation}
\log\left({\frac{\nu}{\nu_B}}\right)={\frac{A_4-0.071x}{0.782-0.545x}},
\end{equation}

\begin{equation}
A_1=-9.30+0.30\delta+(1.30+0.98\delta)\log\left({\frac{\nu}{\nu_p}}\right)
+\log T_{b\nu},
\end{equation}

\begin{equation}
A_2=0.34+0.07\delta,
\end{equation}

\begin{equation}
A_3=0.52+0.08\delta,
\end{equation}

\begin{equation}
A_4=0.10+0.035\delta-\log r_c.
\end{equation}

Here, $x=\cos\theta$, the electron gyrofrequency $\nu_B=2.8\times 10^6 B$.
All coefficients in Eq.(3)-(6) depend on five observable values, i.e.,
radio frequency $\nu$, the turnover frequency $\nu_p$, the brightness
temperature $T_{b\nu}$ at a given frequency $\nu$, the polarization
degree $r_c$, and the electron spectral index $\delta$, which is
calculated by $\delta\approx(1.22-\alpha)/0.9$, here, $\alpha$ is
the photon spectral index in the optically-thin part (Dulk and Marsh 1982).
Thus, it is easy to cancel out the terms with $B$ in Eqs.(1) and (2),
and to obtain a new equation of $x=\cos\theta$ as following.

$$0.782A_1-A_3A_4+(0.071A_3-0.545A_1)x+$$
\begin{equation}
+0.5A_2(0.782-0.545x)\log(1-x^2)=0.
\end{equation}

The theoretical error of the approximations in Dulk and Marsh (1982)
is better than $30\%$ in respect to the full expressions of nonthermal
GS emission (Takakura and Scalise 1970), under the conditions of
$2\leq\delta\leq7$, $\nu/\nu_B\geq10$, the propagation angle
$\theta>20^{\circ}$, the low cutoff energy $E_0=10 keV$, and the
emission is dominated by the extraordinary mode, but the accuracy
worsens at $\delta>6$, especially, at extremes of $\theta$ and
$\nu/\nu_B$ (Dulk 1985).

Now, we pay particular attention to the propagation angle $\theta$
in Eq.(7). When $\theta<90^{\circ}$, it means a positive magnetic
polarity in respect to ordinary and extraordinary mode emission,
together with LCP and RCP sense along the line-of-sight,
respectively. In contrast, when $\theta>90^{\circ}$, it means that a
negative magnetic polarity appears in the radio source, the emission
with LCP sense along the line-of-sight will turn to the X-mode, and
the emission with RCP sense along the line-of-sight will turn to the
O-mode.

However, the polarization is always positive, and the angle $\theta$
is always smaller than 90 degrees in the approximated formulae of
Dulk and Marsh (1982). Actually, the absolute degree of polarization
must be the same at either side of $\theta=90^{\circ}$, but the
sense of polarization is opposite in the two cases, and the
polarization sense changes its sign around $\theta=90^{\circ}$
according to the general plasma dispersion relation. Hence, we
cannot determine the exact direction of the longitudinal magnetic
field with such a 180-degree ambiguity.

With typical values of $T_{b\nu}$, $r_c$, $\nu$, $\nu_p$, and
$\alpha$ in microwave bursts, when $0\leq x\leq 1$ or
$0\leq\theta\leq\pi/2$, we find the left side of Eq.(7) varies
monotonously with $x$ from positive values to negative values, which
means that a unique solution possibly exists for Eq.(7). Figure 1
shows the solution of Eq.(7) with various values of $T_{b\nu}$,
$r_c$, $\nu_p$, and $\alpha$, here, the radio frequency $\nu$ is
fixed as 17 GHz for NoRH. Thus, we can actually minimize the value
of the left side of Eq.(7) to a small value (e.g., $10^{-2}$) to
estimate the solution of $\theta$.

\begin{figure}
\includegraphics[width=8.0 cm]{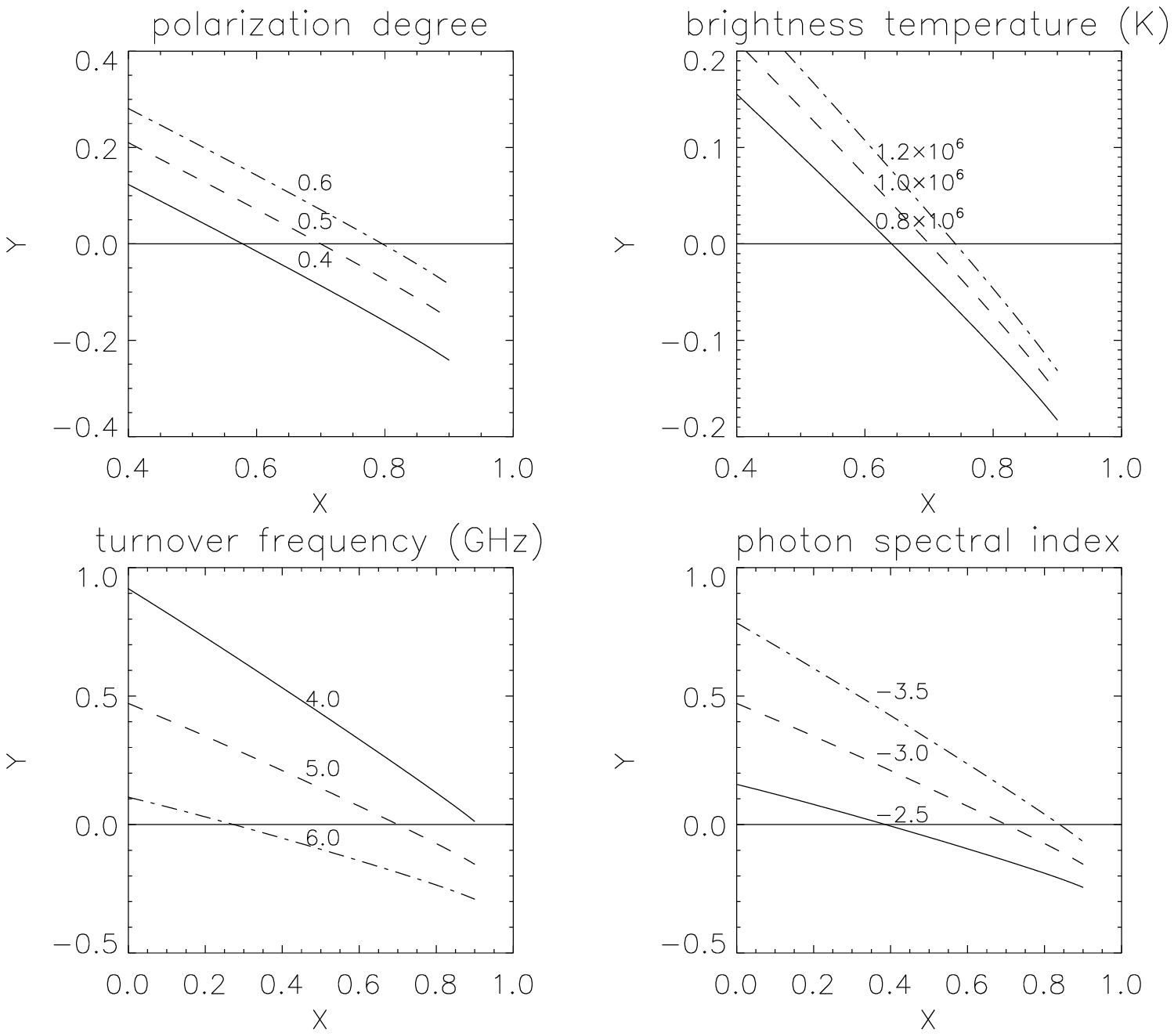} \caption{Left top panel:
the solution of Eq.(7) with different polarization degrees of 0.4
(solid), 0.5 (dashed), and 0.6 (dot-dashed). Left bottom panel: the
solution of Eq.(7) with different brightness temperatures of
$0.8\times 10^6K$ (solid), $1.0\times 10^6K$ (dashed), and
$1.2\times 10^6K$ (dot-dashed). Right top panel: the solution of
Eq.(7) with different turnover frequencies of 4.0 (solid), 5.0
(dashed), and 6.0 GHz (dot-dashed). Right bottom panel: the solution
of Eq.(7) with different photon spectral indices of -2.5 (solid),
-3.0 (dashed), and -3.5 (dot-dashed) The other parameters are always
fixed as the middle value in their varied ranges for all the
panels.}
\end{figure}

In principle, we may determine the intrinsic mode and linear mode
coupling in solar microwave bursts by three steps.

1) At first, the mode coupling may happen only when the angle
approximately equals to 90 degrees, i.e., when the emission
propagates through the quasi-transverse region (Cohen 1960).

2) Secondly, the inversion of polarization sense due to the weakly
mode coupling takes place only when the transition frequency defined
by Cohen (1960) is larger than the frequency of microwave emission,
and an observable criterion of the weakly mode coupling in flaring
loops is indicated by the same polarization sense in the two
footpoints of a flaring loop (Melrose and and Robinson 1994).

3) Finally, the intrinsic mode of microwave emission is determined
by the observed polarization and the calculated direction of local
magnetic field according to the general plasma dispersion relation,
together with the mode coupling process. However, a 180-degree
ambiguity still exists in the direction of longitudinal magnetic
field, to produce an uncertainty of the intrinsic mode.

\section{ANALYSIS OF ONE NORH EVENT}

An M7.6 flare in active region NOAA 9628 (S21E01) start at
04:25:09 UT, peak at 04:34:00 UT, and end at 05:11:47 UT in 2001
September 25, is selected from the flare list in the web of NoRH.
The full Sun observations of Nobeyema Radio Polarimeters (NoRP)
at 1-35 GHz in the same event (Figure 2) are used to discuss
the radiation mechanism of 17 and 34 GHz observed by NoRH.
It includes light curve, polarization, spectrum and turnover
frequency during the event.

\begin{figure}
\includegraphics[width=8.0 cm]{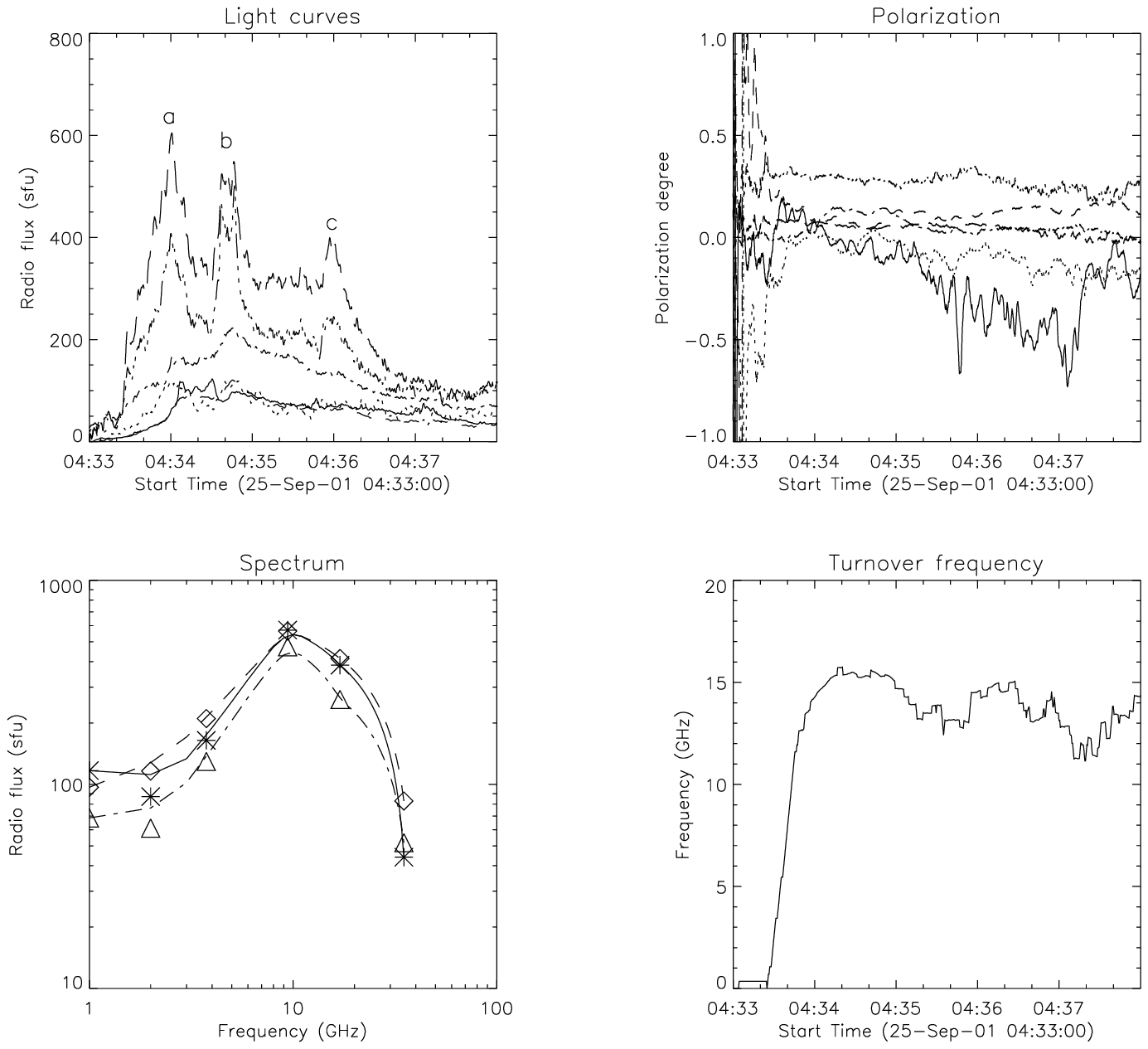} \caption{Left top panel:
Light curves of 1 GHz (solid), 2 GHz (dashed), 3.75 GHz
(dot-dashed), 9.375 GHz (long dashed), 17 GHz (dot-dot-dashed), and
35 GHz (dotted). Right top panel: Polarization degrees of 1 GHz
(solid), 2 GHz (dashed), 3.75 GHz (dot-dashed), 9.375 GHz (long
dashed), 17 GHz (dot-dot-dashed), and 35 GHz (dotted). Left bottom
panel: Spectra at three selected times as marked by 'a' (solid), 'b'
(dashed), and 'c' (dot-dashed) in the left top panel. Right bottom
panel: Time evolution of the turnover frequency.}
\end{figure}

There are different polarization senses in different frequencies,
such as the LCP sense in 1 GHz, and RCP sense at 17 GHz, which is
consistent with the polarization at 17 GHz observed by NoRH as shown
in the left bottom panel of Figure 4. The spectra at three selected
times show the typical features of nonthermal GS emission, but we
still can't exclude the plasma emission that possibly existed at 1-2
GHz (i.e., the CD type with second peak in decimeter band by Nita
{\it et al.} 2004). The turnover frequency increased fast in the
rising phase, then, it varied in 10-15 GHz. Due to larger errors of
peak frequency that fitted from only six frequencies of NoRP, we
suppose that 35 GHz always belongs to optically thin, but 17 GHz is
somehow close to the peak frequency in this event, which may cause
an error in the calculated spectral indices of NoRH.

The left top panel of Figure 3 shows a bright soft X-ray loop
observed by Yohkoh/SXT in this event, overlaid by the contours of
the NoRH/17GHz brightness temperature (solid) and polarization
degree (dashed). The soft X-ray emission in looptop (LT) is much
stronger than almost symmetric emissions in two footpoints (FP1 and
FP2), while, the microwave emissions in the two FPs are very
asymmetric, with a large difference of almost one order of magnitude
during the burst, as shown in the left bottom panel of Figure 3.

\begin{figure}
\includegraphics[width=8.0 cm]{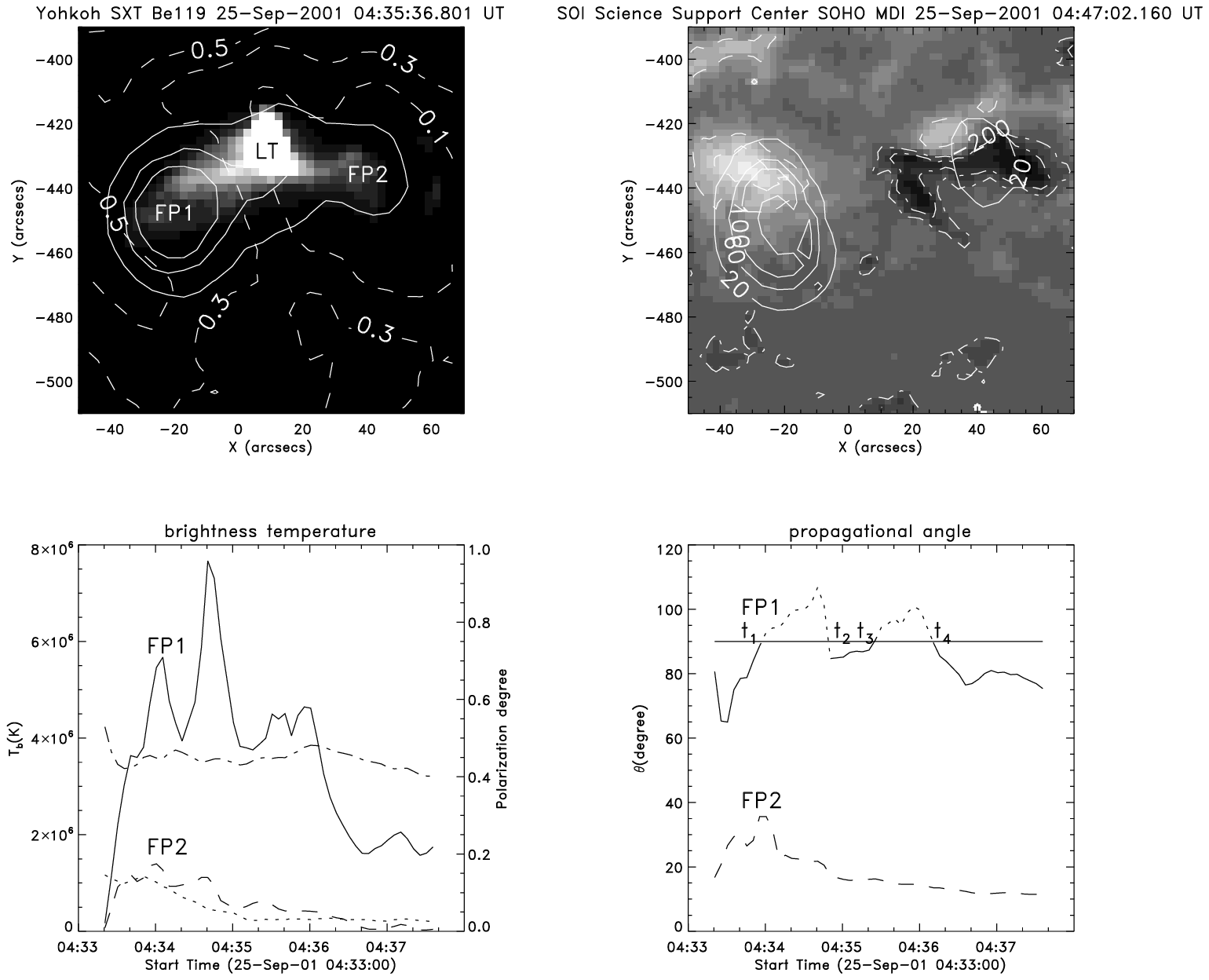} \caption{Left top panel:
Yohkoh/SXT image overlaid by NoRH/17GHz contours of 0.1, 0.3 and 0.5
maximum brightness temperature (solid) and polarization of 0.1, 0.3,
and 0.5 (dashed). Left right panel: MDI magnetogram overlaid by the
calculated propagational angle of 20, 60, and 100 degree (solid),
and magnetic strength of 1500 and 2500 Gauss (dashed), -200 and -800
Gauss (dot-dashed). Left bottom panel: the light curve and
polarization degree of NoRH/17GHz in FP1 (solid and dot-dashed,
respectively) and FP2 (dashed and dotted, respectively), the
position of FP1 and FP2 is marked in the left top panel. Right
bottom panel: the time evolution of the propagational angle in FP1
(solid and dotted) and FP2 (dashed), the horizontal line marks the
angle of 90 degrees.}
\end{figure}

Therefore, we calculate the propagational angle $\theta$ from Eq.(7)
in this event with the brightness temperature ($T_{b\nu}$) at 17 and
34 GHz, and the polarization degree ($r_c$) at 17 GHz observed by
NoRH, and the spectral index ($\alpha$) calculated with the
particular software of NoRH at 17 and 34 GHz with different spatial
resolutions, together with the peak frequency ($\nu_p$) estimated by
NoRP.

It is most interesting that the propagation angle $\theta$ in FP1 is
much larger than that in FP2, and overlaid on the photospheric
magnetogram of MDI in this event (see the right top panel of Figure
3). To confirm this result, we draw the time variation of the
propagation angle $\theta$ in the two FPs in the right bottom panel
of Figure 3. It is evident that $\theta$ reaches $90^{\circ}$ around
the peak time in FP1 (the variation of $\theta>90^{\circ}$ is marked
by dotted line), but $\theta$ in FP2 mostly varies around
$20^{\circ}$. The calculated propagation angles in the two FPs just
mean that there are opposite magnetic polarities in the two FPs
during the impulsive phase of the event, i.e., negative in FP1 and
positive in FP2.

It is predicted that the polarization degree in the optically thin
limit is inversely proportional to the propagation angle from Figure
3d and Eq.(16) of Dulk and Marsh (1982), which seems to conflict
with the results in Figure 4 that shows stronger polarization and
larger propagation angle in FP1, while, weaker polarization and
smaller propagation angle in FP2. The apparent discrepancy between
the theoretical prediction and the results in Figure 4 may be caused
by the asymmetric magnetic fields in the two FPs. The averaged
photosphere magnetic field in FP1 is about 5-10 times larger than
that in FP2, which may have stronger effect on the polarization
degree than that of the propagation angle. The weak polarization
commonly corresponds to the small magnetic field, and the
calculations in Figure 3d of Dulk and Marsh (1982) should be
obtained with the same magnetic field strength.

Moreover, the reason why the apparent photospheric magnetic
polarities in the two FPs are opposite to the calculated coronal
magnetic polarities may be not true due to the 180-degree ambiguity
as mentioned above.

\section{DISCUSSIONS}

In the general plasma dispersion equation, the extraordinary mode
(X-mode) emission along the line-of-sight corresponds to the
observed right-circular polarization (RCP) emitted from a
positive magnetic polarity, and the ordinary mode emission
(O-mode) along the line-of-sight corresponds to the observed
RCP emitted from a negative magnetic polarity. Note that the
left-circular polarization (LCP) is detected as frequently as the
RCP in solar radio observations (see the recent statistics of Huang
and Nakajima 2009). In this case, the X-mode emission along the
line-of-sight may correspond to the observed LCP emitted from
a negative magnetic polarity, and the O-mode emission along the
line-of-sight corresponds to the observed LCP emitted from a
positive magnetic polarity (e.g., Altyntsev {\it et al.} 2008).

The key problem is that we may make mistake by simply using the
photospheric magnetic polarities to determine the intrinsic mode of
the electromagnetic waves detected by solar radio telescopes. If we
use the observed photospheric magnetic polarities to determine the
intrinsic modes in the 2001 September 25 flare, the result seems to
be opposite to that using the calculated longitudinal magnetic
polarities in this event. But, we still have a 180-degree ambiguity
about the direction of longitudinal magnetic field in present case.

Regarding the linear mode coupling, the selected event is very
closed to the central meridian passage (CMP), thus it is difficult
to distinguish the diskward and limbward sources in the two FPs, and
to judge the emission from which FP source propagate through the
QT-region. We can't be simply neglect the mode coupling in the two
FP sources nearby the CMP, which may be caused by a small deviation
of the finite size of the FP sources from the CMP. Moreover, the
observed polarization at 17 GHz in this event has same RCP sense in
both two FPs together with opposite magnetic polarities, which
implies that the transition frequency should be larger than 17 GHz,
i.e., in the weakly coupled case (Melrose and Robinson 1994).

The calculated angle between the local magnetic field and the
line-of-sight in FP1 varied around $90^{\circ}$ at four times
($t_1$, $t_2$, $t_3$ and $t_4$) of the impulsive phase (Figure 3),
which gives a direct evidence of the FP1 emission propagating
through a QT region where the linear mode coupling took place.
While, the calculated angle in another FP was always much smaller
than $90^{\circ}$, which implies that the FP2 emission does not
propagate through the QT region. On the other hand, the observed
photospheric magnetic field in FP1 is about one order of magnitude
larger than that in FP2 (Figure 3), it is ready to compare the the
transition frequency defined by Cohen (1960) and the observed
frequency (17 GHz) of NoRH in both two FP sources, and it is
confirmed that the mode coupling in FP1 actually belongs to the
weakly coupled case. Moreover, the calculated angle between the
line-of-sight and the local magnetic field becomes large with
increasing of radio emissions, which just means an enhancement of
the transverse magnetic field during the burst, and it may support
the existence of the QT-region.

Now, we try to understand the intrinsic mode of the two FP emissions
in this event. From the NoRP data as mentioned above, the 35 GHz
emission is located in the optically-thin region, which belongs to
the X-mode according to Ramaty's theory (1969), while, the 17 GHz
emission is close to the peak frequency and it may belong to X-mode
or O-mode. If the angle $\theta<90^{\circ}$ or the longitudinal
magnetic field is positive, when the 17 GHz emission in FP1
propagates through the QT-region, where the intrinsic mode with the
RCP sense should be reversed from the X-mode to the O-mode in the
two time intervals of the impulsive phase. The reversion from the
X-mode to the O-mode took place at $t_1$ and $t_3$, and the
reversion from the O-mode to the X-mode took place at $t_2$ and
$t_4$. While, the 17 GHz emission in FP2 does not propagate through
the QT region, and the intrinsic mode in FP2 always belongs to the
X-mode with the RCP sense. On the other hand, due to the 180-degree
ambiguity, the angle $\theta$ may equal to $180^{\circ}$ minus the
calculated value. Thus, the intrinsic mode in FP1 is just reversed
from the O-mode to the X-mode at $t_1$ and $t_3$, and the reversion
from the X-mode to the O-mode took place at $t_2$ and $t_4$. While,
the 17 GHz emission in FP2 always belongs to the O-mode.

Finally, there are two results with something new in this paper to
be summarized as follows.

1) According to the theoretical prediction of Melrose and Robinson
(1994), for two FP emissions with opposite magnetic polarities (such
as the event in this paper), they should show the same polarization
sense in the weakly coupled case, but opposite polarization senses
in the strongly coupled case. The problem is that the flare is at
the central meridian, and there should be rough symmetry in the ray
paths to Earth. Hence, the first new result in this paper is just to
give an evidence of the weakly mode coupling near by the central
meridian, which is consistent with the theoretical prediction of
Melrose and Robinson (1994).

2) Secondly, we have derived a new equation of $\cos\theta$ based on
the earlier paper of Huang (2006), and shown the existence of the
solution of this equation under some typical parameters of microwave
bursts, which should be helpful for studying the model coupling and
coronal magnetic field quantitatively. Though the uncertainty of the
sign of $\cos\theta$ does exist, but it does not affect the
prediction of the mode coupling around $\theta=90^{\circ}$.

Regarding the problem of intrinsic modes, we may compare the
extrapolation of photosphere magnetic field with the calculated
coronal magnetic field, and determine the exact direction of coronal
magnetic field, which we plan to do in a forthcoming paper. In this
paper, if we believe that the magnetic polarities in photosphere and
corona are commonly coincident, the intrinsic modes can be
determined therein.

The remained problem that the authors could not solve so far is that
the uncertainty of peak frequencies in different locations. It is
expected to study this problem when the new instruments (such as
FASR and Chinese radio heliograph with multiple frequencies) are
completed in near future.

\begin{acknowledgements}
The study is supported by two NFSC projects respectively with
No.11073058 and 10833007, as well as the "973" program with
No.2011CB811402. The authors would like to thank the NoRH teams
for their dedicated data preparing and preprocessing.
\end{acknowledgements}

\end{document}